\documentclass[a4paper,fleqn]{cas-sc}
\pdfoutput=1 

\usepackage[numbers]{natbib}
\usepackage[T1]{fontenc} 
\usepackage{siunitx}
\usepackage{color}
\usepackage{indentfirst}

\begin{document}
\let\WriteBookmarks\relax
\def\floatpagepagefraction{1}
\def\textpagefraction{.001}
\shorttitle{}
\shortauthors{Y. Tao et~al.}

\title[mode = title]{\boldmath Dark Matter Directionality Detection performance of the Micromegas-based $\mu$TPC-MIMAC detector}

\author[1]{Y. Tao}[orcid=0000-0002-6424-8131]
\cormark[1]
\ead{taoy15@mails.tsinghua.edu.cn}
\author[2]{C. Beaufort}
\author[1,4]{I. Moric}
\author[3,1]{C. Tao}
\author[2]{D. Santos}
\author[2]{N. Sauzet}
\author[2]{C. Couturier}
\author[2]{O. Guillaudin}
\author[2]{J.F. Muraz}
\author[2]{F. Naraghi}
\author[4,1]{N. Zhou}
\author[3]{J. Busto}

\address[1]{Tsinghua Center for Astrophysics, Department of Physics, Tsinghua University, Beijing 100084, China}
\address[2]{Laboratoire de Physique Subatomique et de Cosmologie, Universit Grenoble-Alpes(UGA), CNRS/IN2P3, Institut Polytechnique de Grenoble, 53, rue des Martyrs, Grenoble, France}
\address[3]{Centre de Physique des Particules de Marseille, Aix-Marseille Universit\'e, CNRS/IN2P3, Marseille, France}
\address[4]{INPAC and School of Physics and Astronomy, Shanghai Jiao Tong University, Shanghai Laboratory for Particle Physics and Cosmology, Shanghai 200240, China}

\cortext[cor1]{Corresponding author}

\begin{abstract}
Directional Dark Matter Detection (DDMD) can open a new signature for Weakly Massive Interacting Particles (WIMPs) Dark Matter. The  directional signature provides in addition, an unique way to overcome the neutron and neutrino backgrounds.
In order to get the directional signature, the DDM detectors should be sensitive to low nuclear energy recoils in the keV range and have an angular resolution better than $20^{\circ}$.
We have performed experiments with low energy ($<30\,\mathrm{keV}$) ion beam facilities to measure the angular distribution of nuclear recoil tracks in a MIMAC detector prototype. 
In this paper, we study angular spreads with respect to the electron drift direction ($0^{\circ}$ incident angle) of Fluorine nuclear tracks in this low energy range, and show nuclear recoil angle reconstruction produced by a monoenergetic neutron field experiment. 
We find that a high-gain systematic effect leads to a high angular resolution along the electron drift direction. The measured angular distribution is impacted by diffusion, and space charge or ion feedback effects, which can be corrected for by an asymmetry factor observed in the flash-ADC profile.
The estimated angular resolution of the $0^{\circ}$ incident ion is better than $15^{\circ}$ at $10$ keV kinetic energy and agrees with the simulations within $20$\%.
As it was not possible to inject ions at angles different from zero with respect to the electric drift field, we have performed experiments with monoenergetic neutrons producing nuclear recoils at all angles.
The distributions from the nuclear recoils have been compared with simulated results based on a modified Garfield++ code.  
Our study shows that protons would be a more adapted target than heavier nuclei for DDMD of light WIMPs.
We demonstrate that directional signature from the Galactic halo origin of a Dark Matter WIMP signal is experimentally achievable, with a deep understanding of the operating conditions of a low pressure detector with its diffusion mechanism.
\end{abstract}



\maketitle

\section{Introduction}
\label{sec:intro}

Weakly Interactive Massive Particles (WIMPs) are one of the best motivated Dark Matter candidates. 
Goodman and Witten \cite{GW} suggested that they would interact with detector nuclei with recoil energies in the keV range (see Ref.~\cite{Gui2011}).
Spergel~\cite{Spergel1988} has pointed out that measuring the angular distribution would be an unique signature to confirm the Galactic halo origin of a Dark Matter signal.

Taking the example of a 10 kg CF$_{4}$ $50$ m$^3$ MIMAC detector with a recoil energy range of ($5$, $50$) keV, angular resolution of $10^{\circ}$ and after $3$ years of operation, Billard \emph{et al.}~\cite{Billard2010a} conclude from simulations that even in the presence of a significant background, the detector could set constraints for spin-dependent interactions comparable or better than existing detectors (PICO 2019 \cite{Amole2019}).

Billard \emph{et al.}~\cite{Billard2011} show that with a $100\%$ sense recognition, an angular resolution of $20^{\circ}$ and with no background contamination, this type of detector could reach a $3\sigma$ sensitivity at $90\%$ C.L. down to $10^{-5}$ pb for a WIMP-proton spin dependent cross section.
O'Hare \emph{et al.}~\cite{OHare2015} claim that to discriminate directional signals of light WIMPs from solar neutrinos, an angular resolution of order $30^{\circ}$ or better is requested.

Several projects of Directional Dark Matter Detection (DDMD) are currently being developed \cite{daw11, miuchi07, Santos2010, Ross2014, Deaconu17, CYGNOCollaboration2019}. 
This paper presents a study of the  performance of a MIMAC detector prototype,  in terms of its angular resolution at low nuclear recoil kinetic energies ($6$ to $26$ keV). 
The experimental setup, presented in Section~\ref{sec:experimental}, consists of a MIMAC chamber prototype coupled to an ion beam facility.  
In Section~\ref{sec:recon} we explain how we define and reconstruct the nuclear recoil track direction and discuss the method used to measure the angular resolution.
As pointed out in~\cite{Tao2020}, the MIMAC readouts on the pixelated anodes need to be convolved with the flash ADC asymmetry, and we include this in our analysis of the angular resolution.
We present the final reconstructed angular resolution along the electron drift direction and show that it is below $15^{\circ}$ at an energy as low as $9.32$ keV.
Reconstruction results of larger recoil angles are also presented by analyzing the data of  565 keV neutron experiments.
In Section~\ref{sec:simulations} we compare the results of our measurements with simulations, including a dedicated study of several systematic effects. 
We show that the detection efficiency may have an impact on the angular resolution.
Moreover, there is a strong indication that hydrogen might be the best target for low energy nuclear tracks for directional detection of low mass WIMPs.
In Section~\ref{sec:application}, we revisit the recoil distribution under the Galactic Dark Matter halo model based on the results of Gaia observation, showing that the current finite angular resolution we expect for MIMAC preserves the dipole feature of the WIMP signal.

\section{Experimental Setup and Principle of Operation}
\label{sec:experimental}

The MIMAC detector consists of a matrix of micro-Time Projection Chamber (TPC) (\cite{Sauli1977, Billard2012, Riffard2016}) developed in a collaboration between LPSC (Grenoble) and IRFU (Saclay). 
Each chamber module contains a pixelated bulk Micromegas coupled to fast self-trigger electronics.
In this work, we are using a $10.8\times10.8\times5$ cm$^{3}$ prototype detector~\cite{Tao2020}. 

The optimized working gas is chosen to be a special mixture (called MIMAC gas): 70$\%$ CF$_{4}$ + 28$\%$ CHF$_{3}$ + 2$\%$ iC$_{4}$H$_{10}$, operating at a pressure of $50$ mbar. 
CF$_{4}$ and CHF$_{3}$ provide the main target $^{19}$F for spin dependent Dark Matter detection. 
$^{19}$F is a light odd nucleus, for which the momentum transfer from low mass WIMP elastic scattering is enhanced.
Besides, a fraction of CHF$_{3}$ will effectively reduce the electron drift velocity to about $1/4$ of the pure CF$_{4}$ case.
The isobutane (iC$_{4}$H$_{10}$) helps to increase the gain thanks to its relatively small pairing energy ($23\,\mathrm{eV}$)~\cite{Sharma1998}.

We used the LHI (Ligne exp\'erimentale \`a Haute Intensit\'e) ion beam line~\cite{Tao2020} to generate $^{19}$F$^{+}$ ions with given kinetic energy.
The required species were filtered out, thanks to a high resolution 0.33 T magnetic spectrometer.
The prototype was coupled to the beam line via a \SI{1}{\micro\meter} hole and the ions are thus injected in the direction of the beam line parallel to the drift field in the chamber. 

The high voltages on the grid (or micromesh) and the cathode were set to $-570\,\mathrm{V}$ and $-1320\,\mathrm{V}$ respectively, building up a drift field of $150\,\mathrm{V/cm}$, while the anode was grounded.
Due to the negative voltage applied on the cathode, an extra component of the kinetic energy ($1.32$ keV) must be added to the original one from the ECR ion source.

Part of the kinetic energy of the incident $^{19}$F$^{+}$ ion is released in the detector active volume by ionization.
Primary electron clouds along the physical track generated from ionization drift are subject to diffusion in the drift field. Avalanches take place in the amplification gap of the Micromegas, producing secondary electrons which trigger strips of pixels in the $X$ and $Y$ directions (pitch of \SI{424.3}{\micro\meter})~\cite{iguaz2011}, and are read out by a self-triggered electronics system developed at LPSC~\cite{Couturier2016, Richer2011}.
The $Z$ coordinate of each primary electron is obtained by multiplying the primary electron drift velocity with the relative timing sampling.

The total ionization energy is measured by a charge pre-amplifier coupled to the grid by a flash-ADC.
Both the anode signal and grid charge collection were sampled at $50\,\mathrm{MHz}$ ($20\,\mathrm{ns}$), and events were recorded as coincidence entries one by one.

The gain of the detector coupled to the preamplifier during  the experiment is estimated to be $\mathcal{O}(10^4)$ from the $5.9\,\mathrm{keV}$ peak of the energy calibration source $^{55}$Fe~\cite{Tao2020}.

We have also taken data with $565$ keV monoenegetic neutrons produced by the AMANDE facility~\cite{Gressier2004} giving nuclear recoils at different angles with respect to the drift direction with a $25$ cm drift chamber and MIMAC gas.
The 565 keV neutrons are produced by protons of $2.3$ MeV on a $^7$Li target, which is 1.5 meter away from the cathode of the MIMAC chamber.
In this experiment, the proton beam is in the direction parallel to the electron drift direction.
The Micromegas set up and readout in the neutron experiments is the same as what we used in the LHI experiment.
We also performed another neutron experiment using the MIMAC-IRSN detector ($10\times10\times17.7\,\mathrm{cm}^3$) with $95\%$ of $^4$He + $5\%$ of CO$_2$ gas mixture as a reference experiment (IRSN experiment) for testing our simulation program and reconstruction algorithm.
The MIMAC-IRSN detector is located at $40^\circ$ with respect to the proton beam, which results in the incident neutron energy to be $460$ keV.

\section{3D Reconstruction of Ion Track and Its Direction}
\label{sec:recon}

The $X$-$Y$ 2D positional information is provided by the secondary electrons created by the MIMAC Micromegas avalanche field (see Figure 2 in~\cite{Tao2020}). 
The sampling of the anode every $20$ ns allows the reconstruction of a 3D cloud of primary electrons for each detected event.
In our experiment, we applied a $150\,\mathrm{V/cm}$ electric field, and thus the drift velocity of primary electrons was $V_\text{drift}$ = \SI{22.9}{\micro\meter}/ns, computed by the MAGBOLTZ code~\cite{Biagi1999}. 
After applying a correction on the effective drift velocity, as discussed in another recent work~\cite{Tao2020}, due to space charge effects in the \SI{512}{\micro\meter} gap of the Micromegas detector, we obtain the 3D primary electron cloud for each ion event, reconstruct the direction of the track and estimate the angular resolution.

\subsection{Definition of angular resolution}
\label{sec:definition}

The direction of a recoil track or an incident ion track is modified in the first collision between the injected ion and nuclei in the gas. The important information is the initial ion direction.
However, this ideal information will be washed out by secondary interactions in the drift and avalanche regions.

In order to overcome this challenge, the strategy for reconstructing a track direction is to perform a 3D linear regression fit on the pixelated electron cloud. 
Then we derive the direction of the fitted track with respect to the drift direction ($Z$-axis). 

\begin{figure}[htbp]
    \centering
    \includegraphics[width=0.7\textwidth]{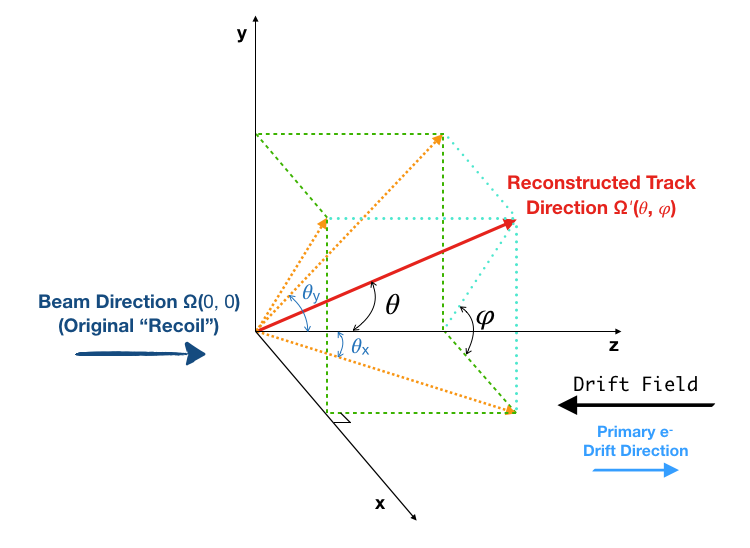}
    \caption{Schematic diagram for direction-related geometrical observables in 3D space. 
    The incoming beam direction is along the $Z$-axis, which is the same as the direction of the drift electric field. 
    An example of reconstructed track direction $\Omega(\theta, \varphi)$ is shown as a red arrow with polar angle $\theta$ and azimuthal angle $\varphi$ indications. 
    The orange arrows represent the 2D projections of this 3D directional vector, defining $\theta_{x}$ and $\theta_{y}$.}
    \label{fig:Coordinate}
\end{figure}

The 3D linear fit on pixelated reconstructed track was performed by a least squares minimizing algorithm using the coordinate distances of the barycenters in each time slice. 
The combination of the straggling and the detector spatial resolution gives the direction of the recoil coming from $\hat{r}(\Omega)$, interpreted as $\hat{r'}(\Omega')$, where $\Omega \equiv \Omega(\theta, \varphi)$ is the solid angle (Figure~\ref{fig:Coordinate}). 

A polar angle $\theta$ was derived for each track, with $0^{\circ}$ being the direction of the $^{19}$F$^{+}$ beam ($Z$-axis and primary electron drift direction). 

$\theta$ is actually the angular deviation from $0^{\circ}$ from all effects combined, after the ion enters the chamber at $0^{\circ}$. 

\begin{figure}[htbp]
  \centering
  \includegraphics[width=\linewidth]{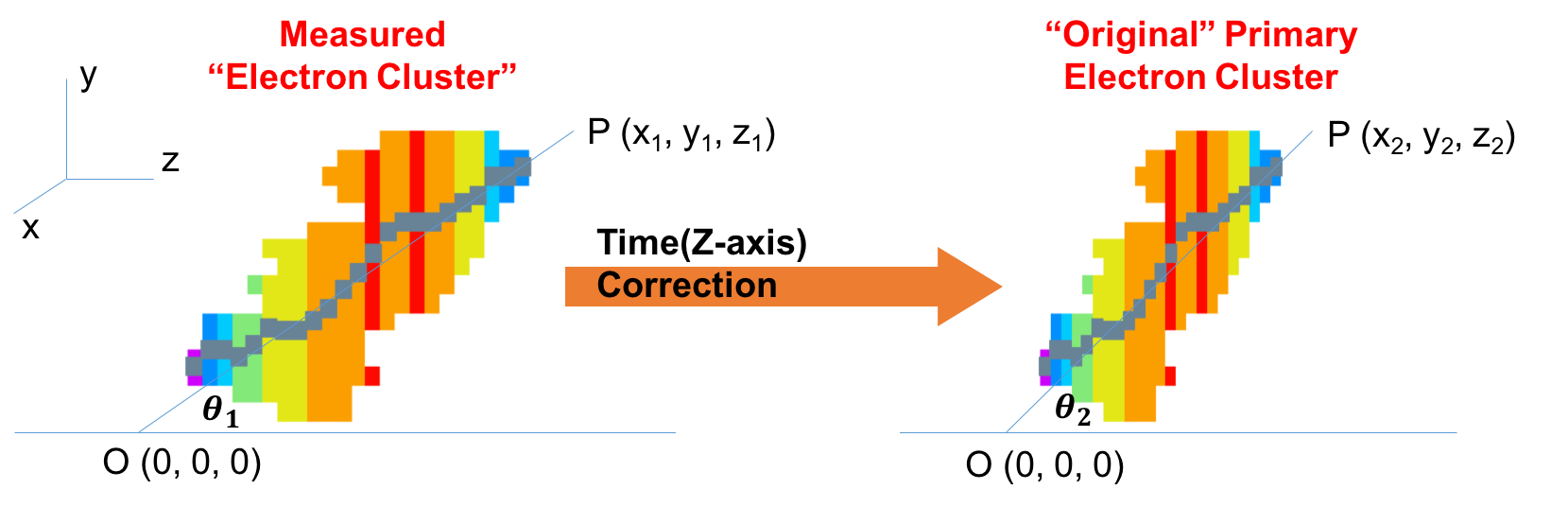}
  \caption{Schematic diagram for obtaining the ``original'' reconstructed direction.}
  \label{fig:angle-corr}
\end{figure}

As discussed later in Section~\ref{subsec:systematics}, systematic effects including the diffusion and space-charge effects can substantially elongate tracks along the $Z$-axis.
In Figure~\ref{fig:angle-corr}, the relationships of spatial coordinates before and after correction are
\begin{equation}
  x_{2}=x_{1}, \quad y_{2}=y_{1}, \quad z_{2}=\eta \cdot z_{1},
\end{equation}
where $\eta$ is the event-by-event correction factor (i.e. \textit{asymmetric factor}).
Constrained by geometry, we have $\tan\theta_{i}=\sqrt{x_{i}^{2}+y_{i}^{2}}/z_{i}, i=1,2$, and then
\begin{equation}
  \theta_{2} = \arctan \left(\frac{z_{1}}{z_{2}} \tan \theta_{1}\right)
  = \arctan \left(\frac{1}{\eta} \cdot \tan \theta_{1}\right).
\end{equation}
$\theta_{2} > \theta_{1}$ for $\eta\in (0, 1]$.
Thus, most ion track will be closer to the initial direction if no high-gain systematic correction is performed.

\begin{figure}[htbp]
    \centering
    \includegraphics[width=0.95\textwidth]{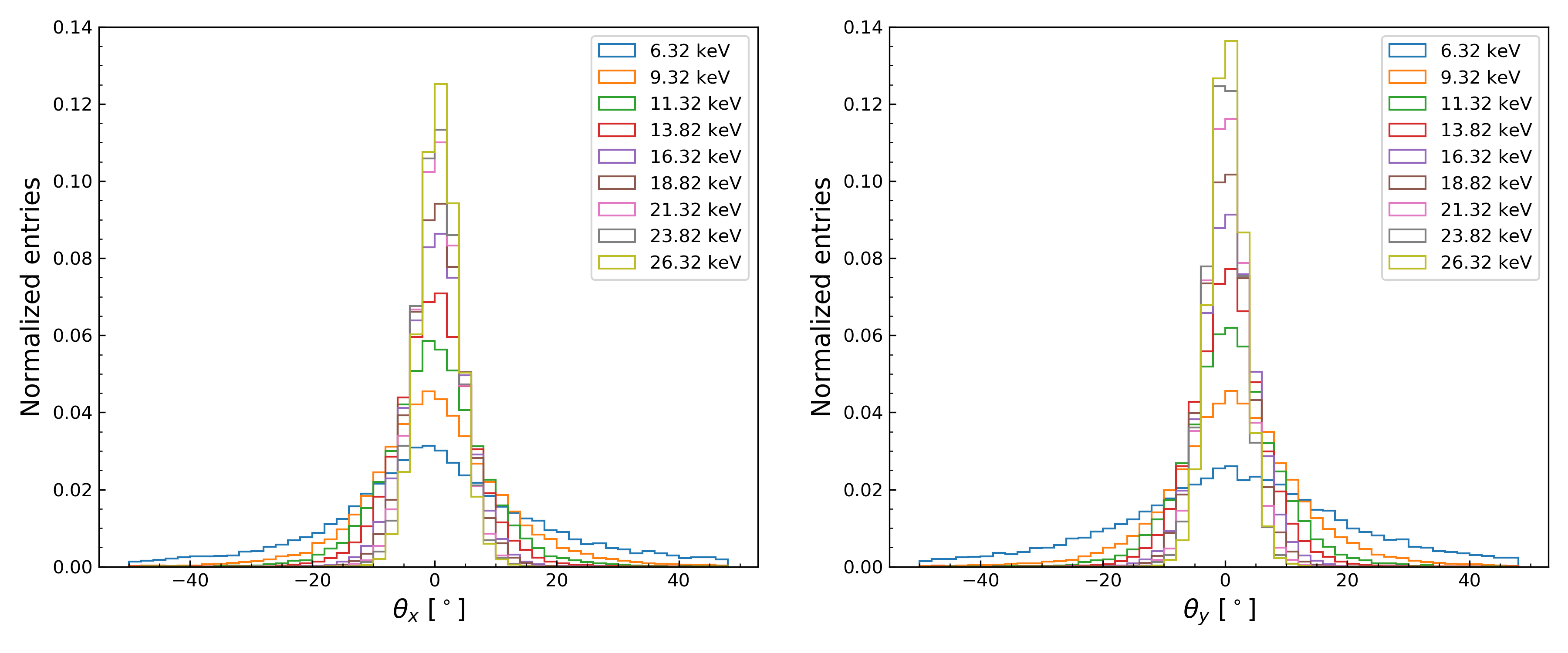}
    \caption{Normalized distributions of $\theta_x$ and $\theta_y$ for $^{19}$F$^{+}$ ions of kinetic energy ranging from 6.3 keV to 26.3 keV.}
    \label{fig:2DTheta}
\end{figure}

The distribution of the reconstructed polar angle can be used to define an angular resolution as discussed below.
However, the distribution of the reconstructed angle $\theta$ between the track and the low energy beam is not a Gaussian variable by definition. 
In contrast, $\theta_x$ and $\theta_y$ defined in Figure~\ref{fig:Coordinate}, appear as Gaussian variables in our experiments as shown in Figure~\ref{fig:2DTheta}. 
All directional information are embedded in $\theta_x$ and $\theta_y$ via
\begin{equation}\label{eq:angle-relation}
    \tan\theta_x = \tan\theta \cos\varphi, \qquad \tan\theta_y = \tan\theta \sin\varphi.
\end{equation}

The spread of angular distribution can be written as 
\begin{equation}\label{eq:resolution_def}
    \sigma_{\theta}(\theta_x, \theta_y)\Big\vert_{\theta_x = \mu_{\theta_x}, \theta_y = \mu_{\theta_y}} 
    = \frac{\sqrt{f^2(\theta_x)\sigma^2_{\theta_x} + f^2(\theta_y)\sigma^2_{\theta_y}}}{f(\theta)},
\end{equation}
in terms of $\theta_x$ and $\theta_y$, where $f(\theta) = (\tan^2{\theta} + 1)\tan\theta$.

Both the angular distribution of the incident ions and the dispersion of the primary electron distribution contribute to the final angular resolution:

\begin{itemize}
    \item Distribution of the incident ions: The reconstructed direction deviates from the initial direction. 
    This is due to several physical effects: (1) primary electrons diffusion, (2) initial ion beam not exactly at zero degree: the hole through which the ions enter the chamber has a \SI{1}{\micro\meter} diameter and \SI{13}{\micro\meter} length (maximum angle of $4.4^{\circ}$) and (3) eventual bias from the reconstruction algorithm.  
    \item Statistical dispersion: Spread of the distribution, usually defined as the standard deviation of a Gaussian Probability Distribution Function (PDF). 
    The main contribution to the statistical dispersion should be the straggling of ions, which is a convolution of multiple small angle scattering with the nuclei of the gas. 
    Other factors deteriorating angular resolution are the interactions of the primary electrons inside the gas chamber, straggling caused by electron collisions and re-combinations with the gas atoms~\cite{Billard2012}, and diffusion~\cite{Peisert84}.
\end{itemize}

The measured mean angle of the incident ions distribution is small ($< 1.6^\circ$) and the dispersion has an effect about 10 times larger than the shift of the central value (more than 4 times for $26.3$ keV), as shown in Table~\ref{tab:GaussParams}. 
Thus we simply take the spread of the angular distribution~(\ref{eq:resolution_def}) as the definition of angular resolution. 

\begin{table}[!htbp]
\centering
\begin{tabular}{|c|c|c|c|c|c|c|c|c|c|}
\hline
$\mathbf{Data\ Label\ (keV)}$ & $\mathbf{6.3}$ & $\mathbf{9.3}$ & $\mathbf{11.3}$ & $\mathbf{13.8}$ & $\mathbf{16.3}$ & $\mathbf{18.8}$ & $\mathbf{21.3}$ & $\mathbf{23.8}$ & $\mathbf{26.3}$ \\
\hline
$\hat{\mu}_{\theta_x}$ & 0.65 & 0.36 & 0.19 & 0.15 & 0.37 & 0.52 & 0.42 & 0.51 & 0.75 \\
\hline
$\hat{\sigma}_{\theta_x}$ & 23.64 & 12.56 & 8.49 & 6.46 & 4.95 & 4.43 & 3.80 & 3.58 & 3.29 \\
\hline
$\hat{\mu}_{\theta_y}$ & 1.37 & 1.58 & 0.85 & 0.51 & 0.51 & 0.11 & 0.09 & -0.09 & 0.30 \\
\hline
$\hat{\sigma}_{\theta_y}$ & 25.25 &12.40 & 8.02 & 6.09 & 4.69 & 4.10 & 3.52 & 3.19 & 2.94 \\
\hline
\end{tabular}
\caption{Gaussian fit parameters of $\theta_x$ and $\theta_y$ distributions for different kinetic energy $^{19}$F$^{+}$ ion events.}
\label{tab:GaussParams}
\end{table}

\subsection{Analysis results on angular resolution}
\label{sec:resolution}

\begin{figure}[htbp]
    \centering
    \includegraphics[width=0.2\textwidth]{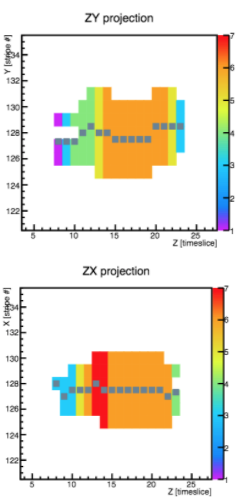}
    \includegraphics[width=0.3\textwidth]{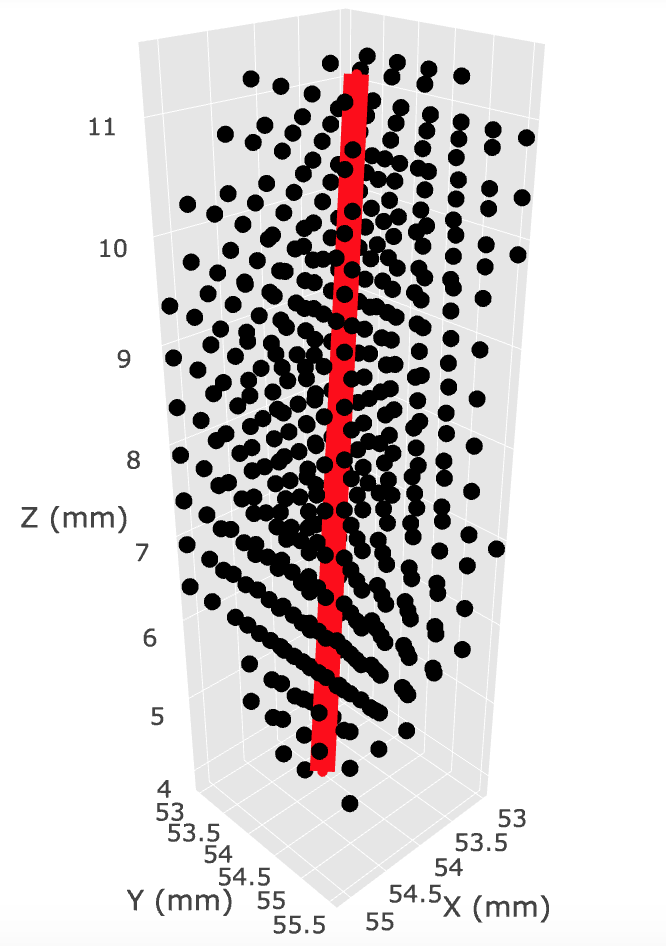}
    \caption{Example of an ion track in $ZX$ and $ZY$ projection using barycenter representation (left) and 3D (right) for an ion of kinetic energy of 26.3 keV. 
    To derive the direction of the track, a 3D linear fit is performed on the 3D cloud of primary electrons.}
    \label{fig:Track26keV}
\end{figure}

\begin{figure}[htbp]
    \centering
    \includegraphics[width=0.2\textwidth]{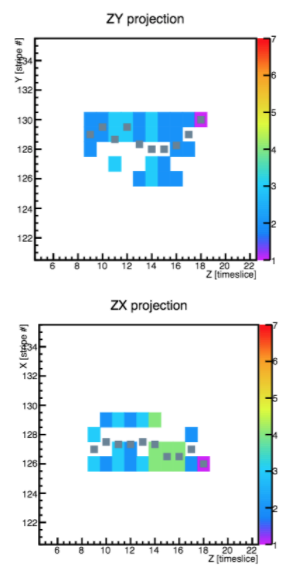}
    \includegraphics[width=0.3\textwidth]{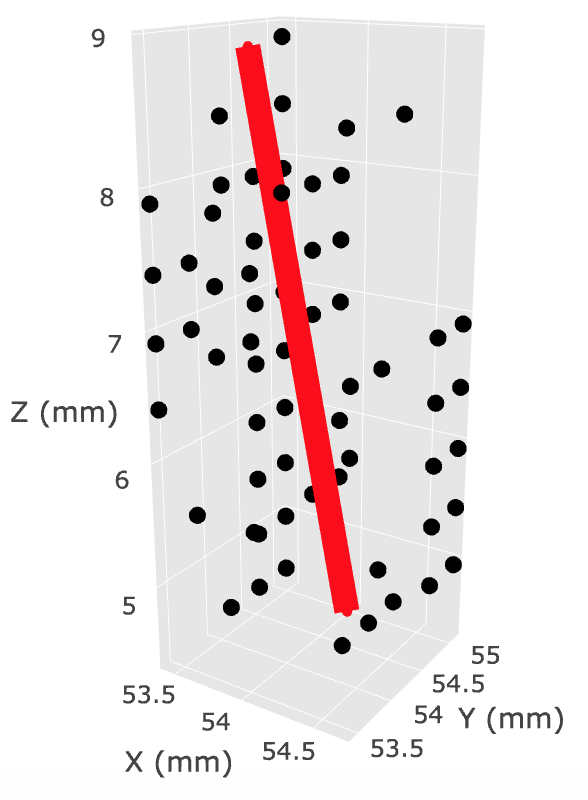}
    \caption{Example of an ion track in $ZX$ and $ZY$ projection using barycenter representation (left) and 3D (right) for an ion of kinetic energy of 9.3 keV. 
    To derive the direction of the track, a 3D linear fit is performed on the 3D cloud of primary electrons.}
    \label{fig:Track9keV}
\end{figure}

The analysis was performed for $^{19}$F$^{+}$ ions with $9$ kinetic energies in the range of $[6.3, 26.3]$ keV, and a statistics of over $1.8\times 10^{4}$ events per each energy, after the background rejection. 
Figure~\ref{fig:Track26keV} and Figure~\ref{fig:Track9keV} show examples of track trajectories in $ZX$, $ZY$ projections for ions with kinetic energies of $26.3$ keV ($25$ keV from the voltage acceleration in ion source plus $1.32$ keV from the cathode voltage with respect to the ground) and $9.3$ keV respectively, with the best fit line in 3D.

\begin{figure}[htbp]
    \centering
    \includegraphics[width=0.9\textwidth]{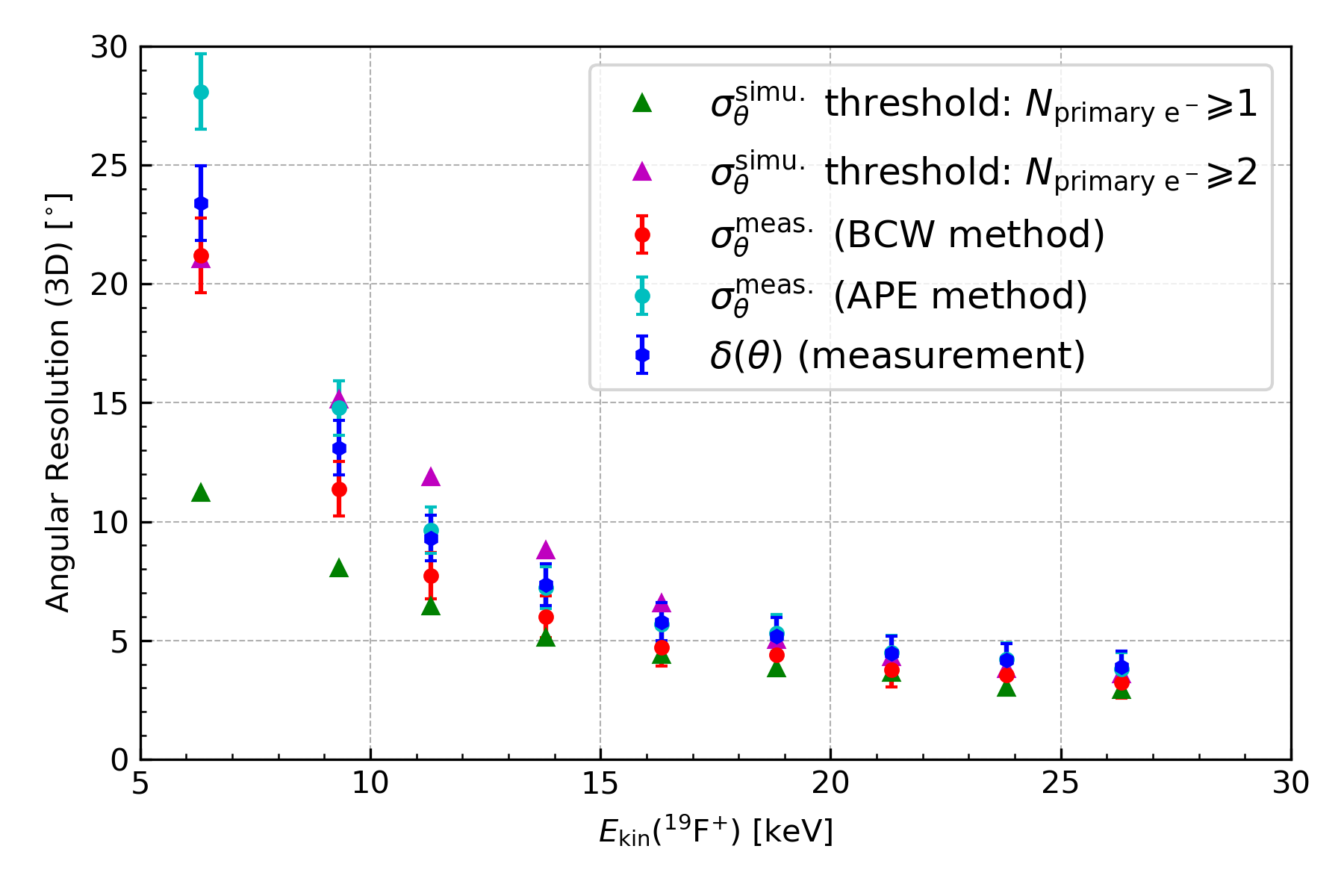}
    \caption{The measurement (red and cyan, with different reconstruction method) and simulation (green and magenta, with different threshold) of angular resolution of MIMAC detector as a function of $^{19}$F$^{+}$ ion kinetic energy. 
    At lower energies, the ion tracks are shorter and have more straggling, resulting in worse angular resolution  (bigger error bars). 
    The measured angular resolution is better than $15^{\circ}$ down to a kinetic energy of $9.3$ keV. 
    Error bars are derived from the pixel strips pitch and reconstructed track length as described in the text.
    We also show the measured mean angle $\delta(\theta)$ between the incident and reconstructed direction for each ion kinetic energy (blue).}
    \label{fig:ETheta}
\end{figure}

The final angular resolution as a function of the ion kinetic energy is shown in Figure~\ref{fig:ETheta} (red and cyan). 
Its dispersion is better than the required $20^{\circ}$ \cite{Billard2011} down to kinetic energy of $9.3$ keV.
We also plot the mean angle between the incident and reconstructed direction (denoted as $\delta(\theta)$) for each ion kinetic energy (blue).

The derived uncertainty (denoted as $\Delta\theta$) on angular resolution is based on the determination of the spatial coordinates of the reconstructed primary electron cloud and the error of the 3D linear fit: 

\begin{equation}\begin{split}\label{eq:errorsys}
    \Delta\theta(x,y,z)\Big|_{\theta = \bar{\theta}} &= \sqrt{(\frac{\partial \theta}{\partial x})^2\Delta^2(x) + (\frac{\partial \theta}{\partial y})^2\Delta^2(y) + (\frac{\partial \theta}{\partial z})^2\Delta^2(z) + \Delta_{\textrm{fit}}^2} \\ &\simeq \frac{\cos^2{\theta}}{z}\sqrt{\Delta^2_{XY} + \tan^2{\theta}\cdot\Delta^2(z)}
\end{split}\end{equation}

where $\Delta(z) = \Delta(V_\text{drift}\cdot t)$ mainly depends on sampling time, 
$\Delta_{XY} = \Delta(x) = \Delta(y)$ is the intrinsic systematic uncertainty due to alignment and finite size of anode strips. 
The fit error $\Delta_{\textrm{fit}}$ is negligible, so we can only take the first term into consideration. 
For $\theta \approx 0^{\circ}$ case, the uncertainty can be further approximated and simplified to be only dependent on the pitch of the anode strips and the reconstructed track length:

\begin{equation}
    \Delta\theta(x,y,z)\Big|_{\theta = \bar{\theta} \simeq 0^{\circ}} = \frac{\Delta_{XY}}{L},
\end{equation}

where $\Delta_{XY}$ is the same as in \eqref{eq:errorsys} and $L$ describes the primary electron cloud dimensions (the reconstructed ion track length after the empirical correction~\cite{Tao2020}).
For the accuracy of our estimation, the error bars presented in Figure~\ref{fig:ETheta} are derived from \eqref{eq:errorsys}.
The error we obtained is $\pm 1.57^{\circ}$ for the lowest ion kinetic energy and $\pm 0.67^{\circ}$ for the highest. 

We have applied various algorithms in order to find the best way to reconstruct the initial angle. 
The differences among these algorithms are mainly whether to use the entire electron cloud or only part of it, and how to set weight on each pixel. 
Modifying the algorithms to use only the first part of the track (with a $\chi^2$ test to select the optimum number of points) does not yield an improvement on the angular resolution. 
In addition, initial and final time slices of the track usually have a larger than average deviation from the track direction. 
This is because the anode samples the endpoints of the transversely diffused primary electron cloud. 
Removing the first and last time slice does not produce better results either.

In this paper, we present results of a barycenter weighted (BCW) reconstruction method and an all pixel equal weighted (APE) reconstruction method.
The average reconstructed angle using BCW method is closest to the initial 0$^{\circ}$ angle, and has the lowest dispersion, while the feasibility of APE method is more general.

\begin{figure}[htbp]
    \centering
    \includegraphics[width=0.95\textwidth]{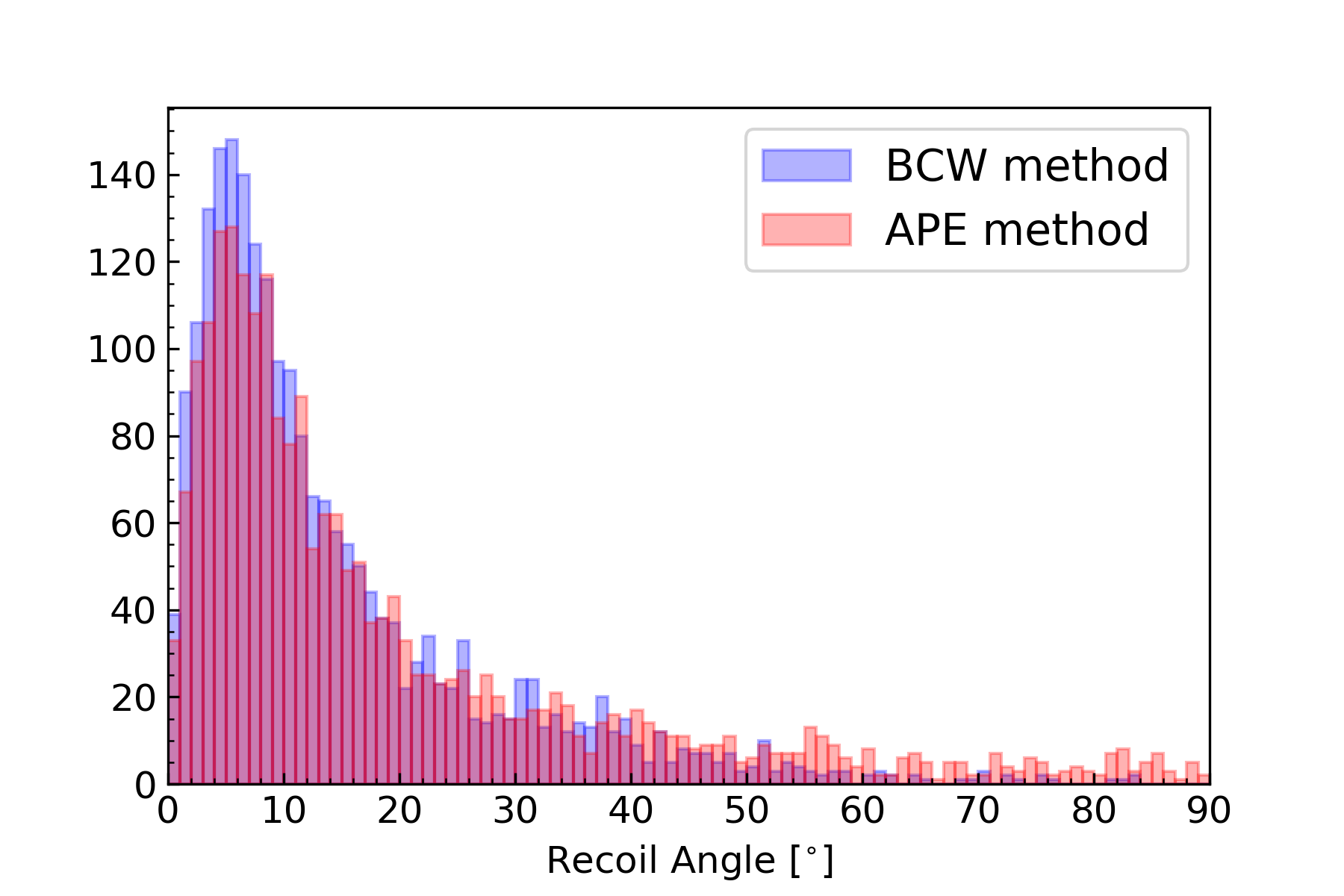}
    \caption{Recoil angle of all recoil tracks in the 565 keV neutron experiment.}
    \label{fig:neutron-expe-recoil}
\end{figure}

In the neutron experiments with the MIMAC gaz, more possible species are scattered by the neutrons and produce recoils, including  proton, Carbon and Fluorine nuclei, and even heavier species CF$_x$.
The reconstructed recoil angles based on BCW and APE methods are shown in Figure~\ref{fig:neutron-expe-recoil}, in the range of [0$^\circ$, 90$^\circ$].

\section{Comparison of simulations with measurements}\label{sec:simulations}

The angular resolution measurements have been compared to a version of the Garfield++~\cite{Veenhof1998} simulations. 
Garfield++ is a toolkit for particle tracking simulations, which proposes an interface to SRIM~\cite{Ziegler2010} to generate ion tracks, and whose electron transport algorithm solves the second-order equations of motion based on the MAGBOLTZ~\cite{Biagi1999} gas tables.

The simulation code has been verified by reproducing the neutron experiment at 27 keV~\cite{Maire2015}, and in this work we adjust it to reproduce the LHI experiment conditions. 
For each kinetic energy, we have sent 200 Fluorine ions at $(X, Y, Z) = (5.4\,\mathrm{cm}, 5.4\,\mathrm{cm}, 5\,\mathrm{cm})$ along the $Z-$direction. 
Each primary electron is transported up to the grid and suffers the transverse and longitudinal diffusion on the way. 

We have observed a continuous reduction of the drift velocity depending on the detector gain and the charge density.  A possible explanation for such a phenomenon is a slow ion backflow in the Micromegas  gap, which builds up space-charge and locally distorts the electric field in the higher gain configuration.
The asymmetry of the flash-ADC signal gives a direct event by event observation of this velocity reduction, which enables us to determine an \textit{asymmetric factor} $\eta$ that quantifies the magnitude of the elongation. 
We have experimentally validated the ability of this factor to describe the phenomenon as a first approximation~\cite{Tao2020}. 
The space-charge effect has been included in the simulations: the drift velocity is linearly reduced according to the value of $\eta$.

The simulations sample the arriving of charges \textit{MIMAC-wise}: a $50\,\mathrm{MHz}$ ($20\,\mathrm{ns}$) readout and a pixelization in the $X$-$Y$ plane with strips at \SI{424.3}{\micro\meter} pitch. 
We  apply the same analysis algorithms on the measurements and the simulations, allowing us to compare the results.

\subsection{Angular resolution along the electron drift direction}

\begin{figure}[htbp]
    \centering
    \includegraphics[width=\linewidth]{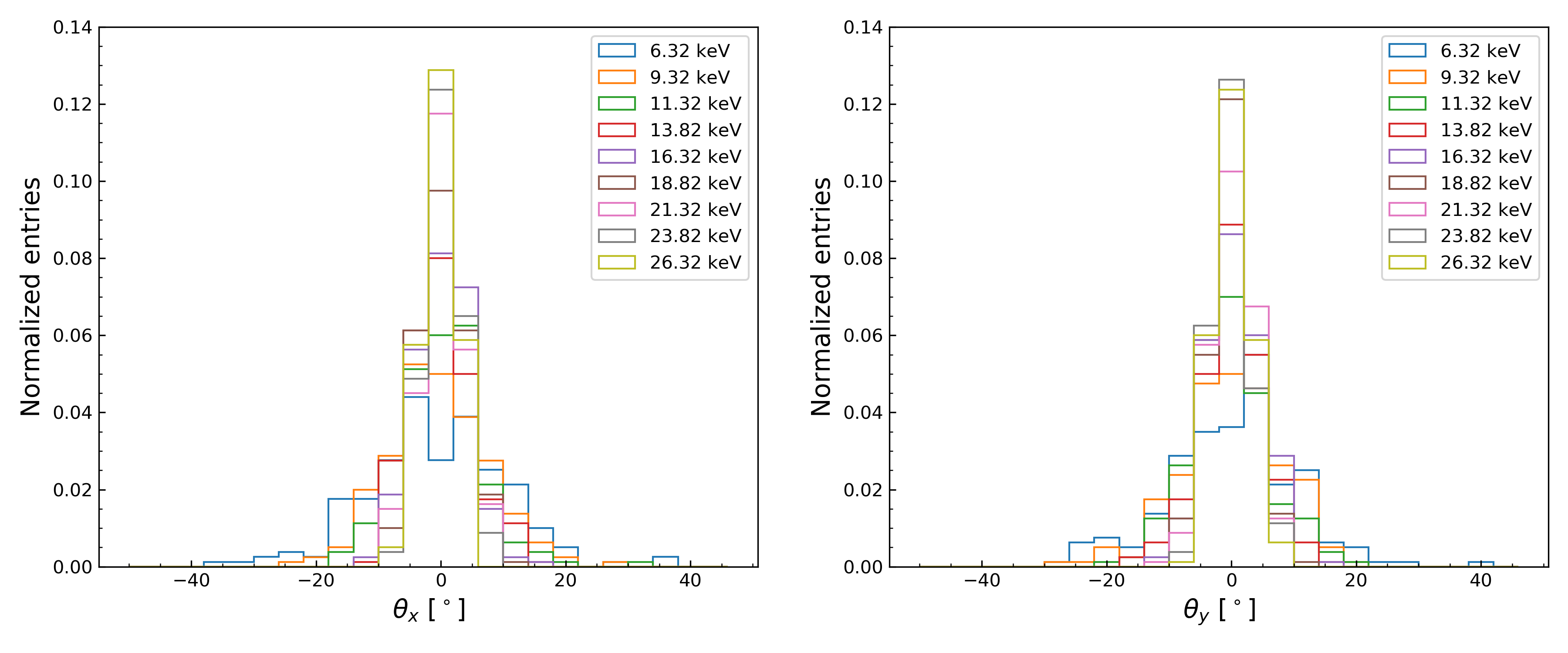}
    \caption{Normalized distributions of the projected angles from Garfield++ simulations with 200 $^{19}$F$^+$ ions per energy.}
    \label{fig:thetaXY}
\end{figure}

The simulation results of the projected angles $\theta_x$ and $\theta_y$ are presented in Figure \ref{fig:thetaXY}. 
The angle dependence on the kinetic energy follows the same evolution observed in the measurements. 
At $6.32\,\mathrm{keV}$, the spread of the simulated distribution is twice narrower than the measured one; this difference decreases with the energy and remains below $20\%$ for energies above $13.82\,\mathrm{keV}$. 
This difference at low energies propagates to the angular resolution, as shown in Figure~\ref{fig:ETheta}. 
An explanation for these differences is given below.

\subsection{Study of systematic effects}
\label{subsec:systematics}

Along its way to the anode, the primary electron cloud suffers several distortions modifying its shape. 
The angular resolution depends not only on the conservation of the primary cloud shape, but also on the experimental access to this information.
The question is: will the physical information of the initial direction be washed out by these shape distortions? 

We identified four main effects: 
\begin{itemize}
  \item [(1)] the diffusion in the drift volume, 
  \item [(2)] the space-charge effect in the amplification region, 
  \item [(3)] the avalanche, 
  \item [(4)] the finite threshold of the anode strips. 
\end{itemize}

We have studied how the angular resolution is affected by each one of these effects.

According to SRIM and without the distortions, a $^{19}$F$^+$ ion track of kinetic energy below $20\,\mathrm{keV}$ would be seen as a few pixels cloud but only one or at most two  time slices of $20\,\mathrm{ns}$ would be triggered in a MIMAC detector chamber and no clear direction would be measured.

We have shown that the diffusion in $5\,\mathrm{cm}$ drift enlarges the primary cloud of about one order of magnitude in each direction ~\cite{Tao2020}.
If the diffusion was fully symmetric, the directional information would be conserved during the cloud enlargement. 
However, according to MAGBOLTZ, the longitudinal diffusion dominates the transverse one by a factor $\frac{293.9}{253.1} = 1.16$.
With the correction for this factor on the polar angle of each track, the diffusion could appear as a helping process for 3D detection of low energy nuclear recoils. 

In addition, the space-charge or ion backflow systematic effect in the avalanche gap reduces the primary charge collection drift velocity giving a reduced total effective drift velocity, and thus more time slices describing the primary electron cloud along the $Z-$direction. 

\begin{figure}[htbp]
    \centering
    \includegraphics[width=0.9\textwidth]{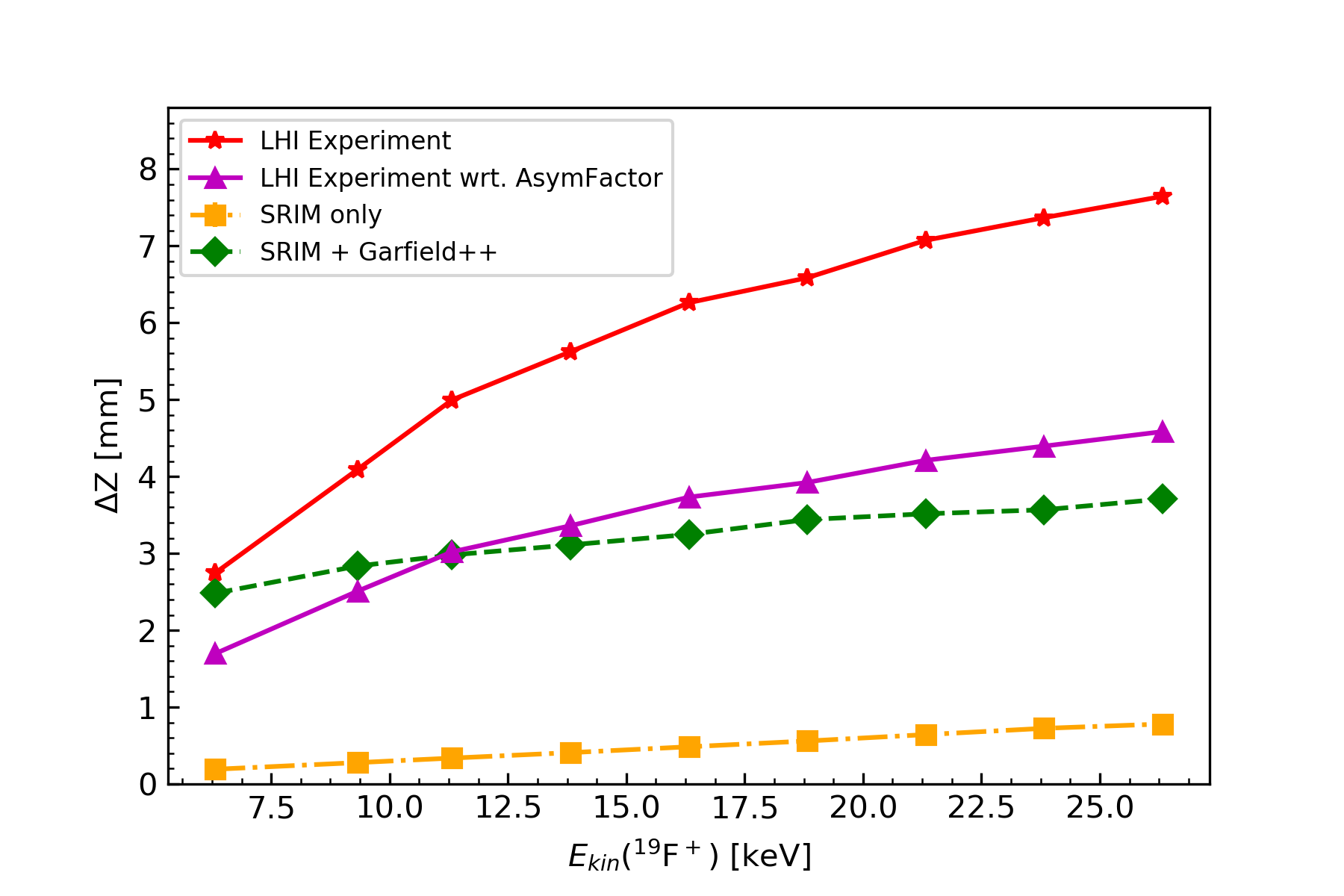}
    \caption{Main results of a companion work~\cite{Tao2020} showing comparison of ion depths ($\Delta Z$) at different energies between experiment (red stars) and Monte Carlo simulation (blue circles) combining SRIM and diffusion.
    The orange box is for SRIM only, the green diamond when diffusion and other effects are included using Garfield++. 
    The magenta triangles are experimental measurements with an asymmetric factor correction.}
    \label{fig:comp-result}
\end{figure}

We have mentioned in a previous work~\cite{Tao2020} ( its Figure 5 is reproduced in Figure~\ref{fig:comp-result} here), that this effect can be described by an \textit{asymmetric factor} $\eta$ between the flash-ADC rising and falling times. 
This multiplicative factor comes from the reduction of the drift velocity under space-charge effects, and has a value in the range $[0.66, 0.71]$ in the conditions of our LHI experiment.
We confirm the space-charge (or ion backflow) systematic effect experimentally with the help of experiments performed with the COMIMAC ion facility, also developed in LPSC (Grenoble) for keV ion researches ~\cite{Tao2020}.

\begin{figure}[htbp]
    \centering
    \includegraphics[width=\linewidth]{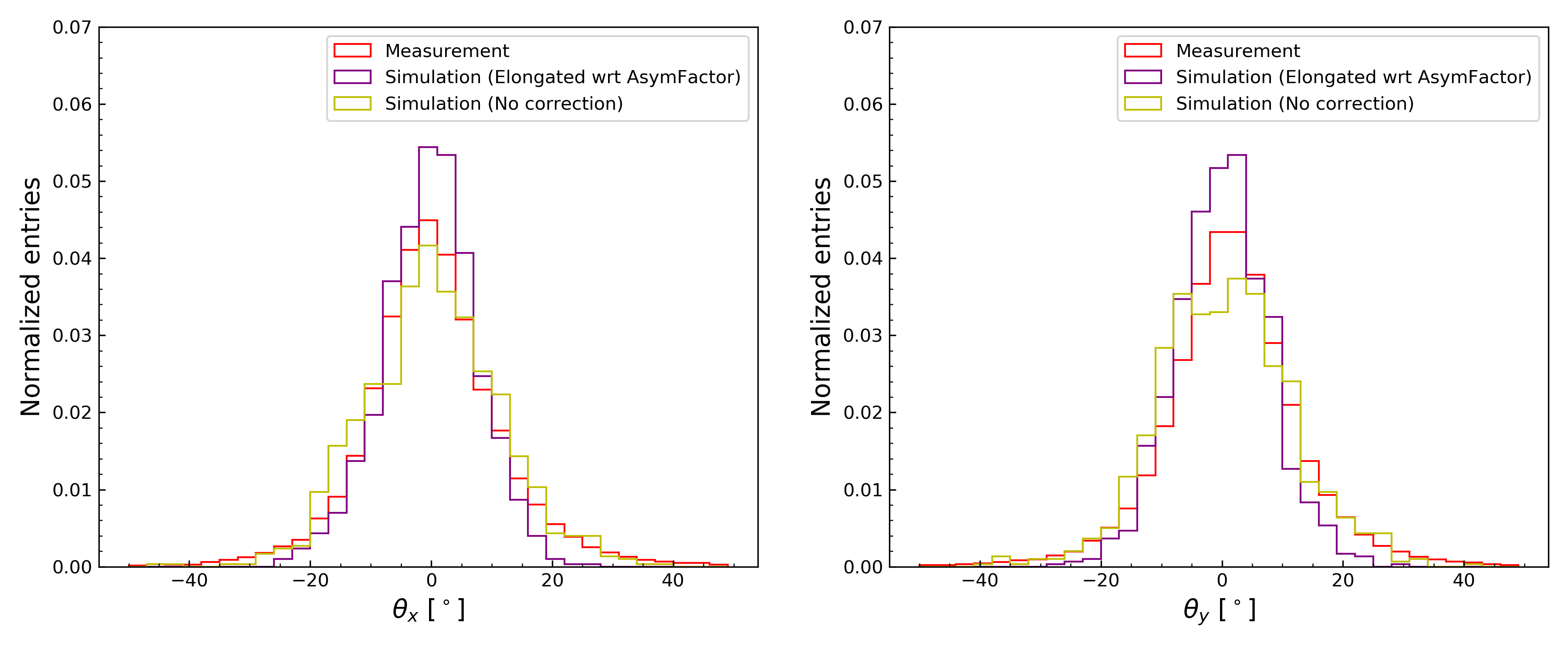}
    \caption{High-gain systematic influence on the projected angles from Garfield++ simulations with 1000 $^{19}$F$^+$ ions of $9.32\,\mathrm{keV}$.
    Measurement results of LHI experiment are also presented for comparison.}
    \label{fig:spaceCharge}
\end{figure}

The Garfield++ simulations give the possibility to include or not this drift velocity reduction in order to study the space-charge effect. 
Figure~\ref{fig:spaceCharge} shows the simulation of $\theta_x$ and $\theta_y$ angles computed from $1000$ $^{19}$F$^+$ ions of $9.32\,\mathrm{keV}$ kinetic energy, with and without the drift velocity reduction. 
The charges are collected in $9$ time slices for the simulated clouds with the reduction; in $6$ time slices otherwise. 
This distortion leads to an underestimated reconstructed angle since the drift velocity reduction deforms the electron cloud towards the electric field drift direction, $0^\circ$ in our case. 
For $9.32$ keV ions, we obtain a better angular resolution when the velocity reduction is corrected for ($8.00^\circ$ vs $11.2^\circ$). 
Note that this resolution improvement is only effective close to $0^\circ$.  
The same reasoning can be conducted for the asymmetric diffusion (stronger along the electric field direction), which has an influence $3$ to $4$ times smaller than the influence of the space-charge effect.

We have shown that the avalanche contribution to the depth was negligible, see Figure~\ref{fig:comp-result}, being symmetric in $X$ and $Y$ directions. 
For this reason, the barycenter weighted method used for the 3D linear fit is not affected by the avalanche enlargement.

Finally, after the avalanche, the positions in the $X$-$Y$ plane are measured by the anode strips. 
Each strip has its intrinsic noise threshold and can only be triggered if the number of charges within one time slice exceeds this threshold. 
This eventual lack of efficiency results in some non-detected charges, especially at low energies. 
We have studied the angular resolution dependence on the strip threshold with Garfield++. 
Since we do not simulate the avalanche, the threshold value is applied on the number of primary electrons needed.
The results are presented in Figure~\ref{fig:ETheta}, where the same threshold value is applied on $X$ and $Y$ strips.

The anode lack of efficiency represents a significant systematic effect, especially at low energies (Figure~\ref{fig:ETheta}).  
Comparison with simulations suggest that the LHI measurements are better described by a threshold of 2 primary electrons than one single primary electron.
	
The Garfield++ simulations have allowed us to isolate each systematic effect in order to study its influence on the angular resolution. 
At low energy, the number of readouts acts as a critical parameter for the 3D linear regression accuracy. 
For this reason, even if the diffusion distorts the primary electron cloud, it appears as a necessary process in order to trigger enough pixels above threshold, and provide a clear detection of sub-millimeter tracks.
The fully asymmetric distortion from the space-charge effect results in an underestimation of the reconstructed angle and consequently becomes a systematic bias in the measurements.

Due to the limited length of low kinetic energy Fluorine ion tracks (similar for Carbon ions), the current direction reconstruction algorithms only work for small angles ($< 20^\circ$), and thus we would need to develop a better adapted reconstruction algorithm to treat them.

\begin{figure}[htbp]
    \centering
    \includegraphics[width=\linewidth]{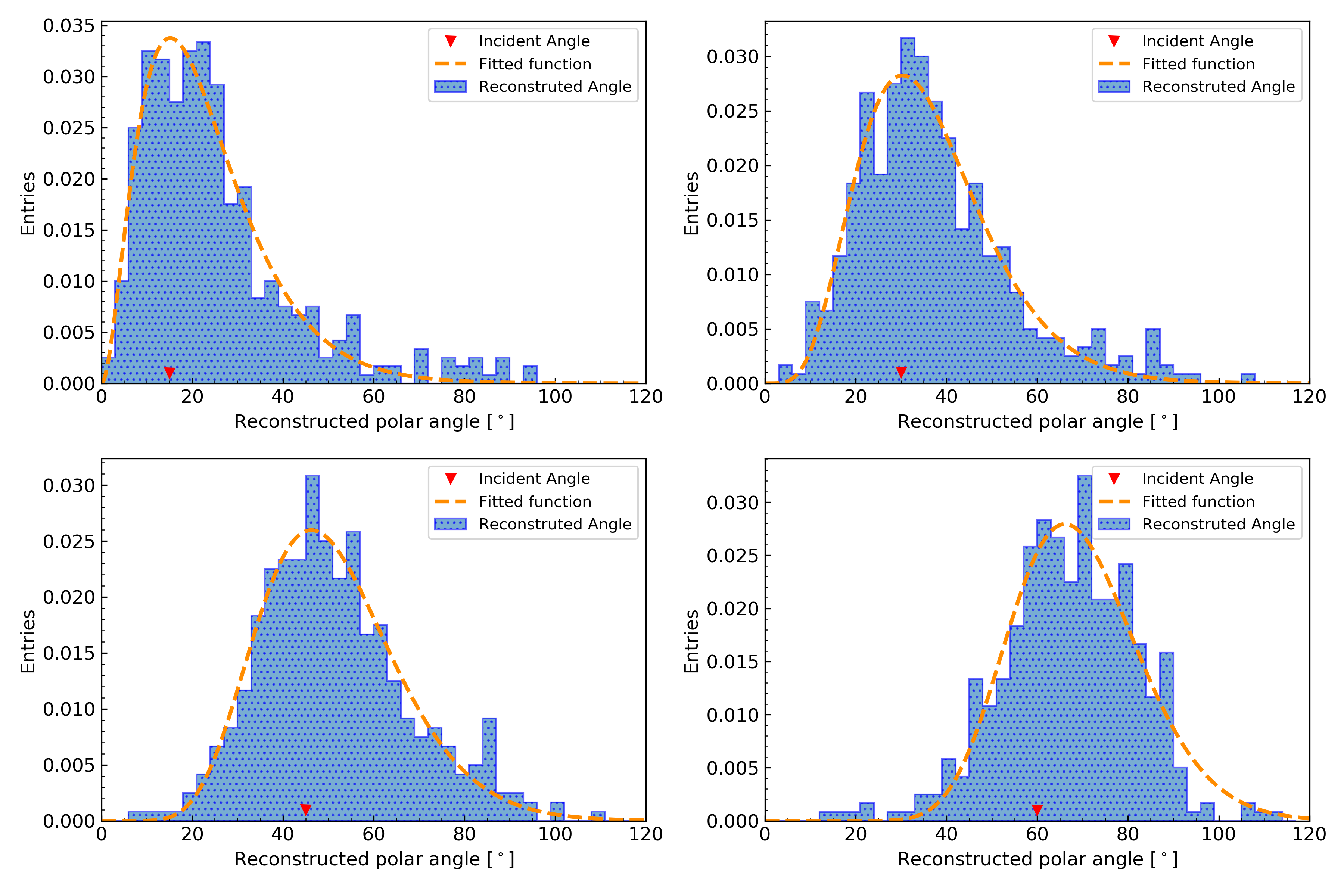}
    \caption{Reconstructed polar angle distribution for 400 incident $21.3$ keV protons with preset directions $15^\circ$, $30^\circ$, $45^\circ$ and $60^\circ$ from Garfield++ simulations with MIMAC gas at $50$ mbar.
    The red inverted triangles label the preset incident polar angles in these 4 cases respectively.
    The orange dashed curve in each subplot is an empirical fit.}
    \label{fig:proton-simu-recon}
\end{figure}

However, the direction for keV hydrogen nuclear tracks (ie. proton tracks) can be well-reconstructed by 3D regression fit, thanks to their longer tracks (compared with heavier nuclei).
Figure~\ref{fig:proton-simu-recon} present the reconstructed polar angle distributions for different initial preset incident directions of proton ($15^\circ$, $30^\circ$, $45^\circ$, $60^\circ$), generated by our Garfield++ simulation.

The reconstructed results are then presented by the most probable value of the empirical fit function and the asymmetric statistical error is computed from the FWHM interval of the polar angle distribution, as shown in Table~\ref{tbl:proton-recon}.
\begin{table}[htbp]
  \centering
  \begin{tabular}{ccccc}
    \toprule[1.5pt]
      Incident polar angle [$^\circ$] & $15$ & $30$ & $45$ & $60$ \\
    \midrule[1pt]
      Reconstructed polar angle [$^\circ$] & $15.14_{-0.43}^{+0.74}$ & $30.27_{-0.59}^{+0.64}$ & $46.25_{-0.63}^{+0.90}$ & $66.07_{-0.61}^{+0.88}$ \\
    \bottomrule[1.5pt]
  \end{tabular}
  \caption{Polar angle reconstruction results for proton tracks, generated by Garfield++.
  The reconstructed polar angle results are the most probable values of the empirical fit function.}
  \label{tbl:proton-recon}
\end{table}

Hydrogen is an odd nucleus which can also probe the spin-dependent WIMP-nucleon scattering process, turns out to be a better target for directional detection of light WIMPs.

\section{Directional Signatures after Gaia Observation}
\label{sec:application}

L. Necib \textit{et al.}~\cite{Necib2018} used the latest data released by Gaia satellite to infer the Dark Matter distribution in the vicinity of 8 kpc from the center of the Galaxy, which is around the Solar system. 
They traced the Galactic Dark Matter by stars and found that the Milky Way halo has two major components, 
\begin{equation}
  f(\boldsymbol{v}) = \xi_\mathrm{h} f_\mathrm{h}(\boldsymbol{v}) + \xi_{\mathrm{s}} f_{\mathrm{s}}(\boldsymbol{v}),
\end{equation}
where $\xi_\mathrm{h,s}$ is the ratio factor that quantifies each component satisfying
$\xi_\mathrm{h} + \xi_\mathrm{s} = 1$.

One is the three-dimensional tri-axial Gaussian distribution halo component, and the other is a substructure component that could be formed by a younger merger event.
The latter component can be characterized by a double Gaussian, which only takes opposite parameters in the radial direction, and other parameters are the same.
\begin{equation}
  \begin{array}{l}
    f_{\mathrm{h}}(\boldsymbol{v}) \propto \mathcal{N}\left(\boldsymbol{\mu}_{\mathrm{h}}, \boldsymbol{\Sigma}_{\mathrm{h}}\right), \quad
    f_{\mathrm{s}}(\boldsymbol{v}) \propto \frac{1}{2}\left[\mathcal{N}\left(-\boldsymbol{\mu}_{\mathrm{s}}, \boldsymbol{\Sigma}_{\mathrm{s}}\right)+\mathcal{N}\left(\boldsymbol{\mu}_{\mathrm{s}}, \boldsymbol{\Sigma}_{\mathrm{s}}\right)\right].
  \end{array}
\end{equation}
Here $\mathcal{N}(\boldsymbol{\mu}, \boldsymbol{\Sigma})$ is the three-dimensional Gaussian distribution with $\boldsymbol{\mu} = (\mu_r, \mu_\theta, \mu_\phi)$ and $\boldsymbol{\Sigma} = (\sigma_r, \sigma_\theta, \sigma_\phi)$ taken from Ref.~\cite{Necib2018}.
With a large uncertainty of its fraction contribution, we can parametrize it with
\begin{equation}
  r = \frac{\xi_{\mathrm{s}}}{\xi_{\mathrm{h}}+\xi_{\mathrm{s}}} \in[0,1].
\end{equation}

\begin{figure}[htbp]
    \centering
    \includegraphics[width=\linewidth]{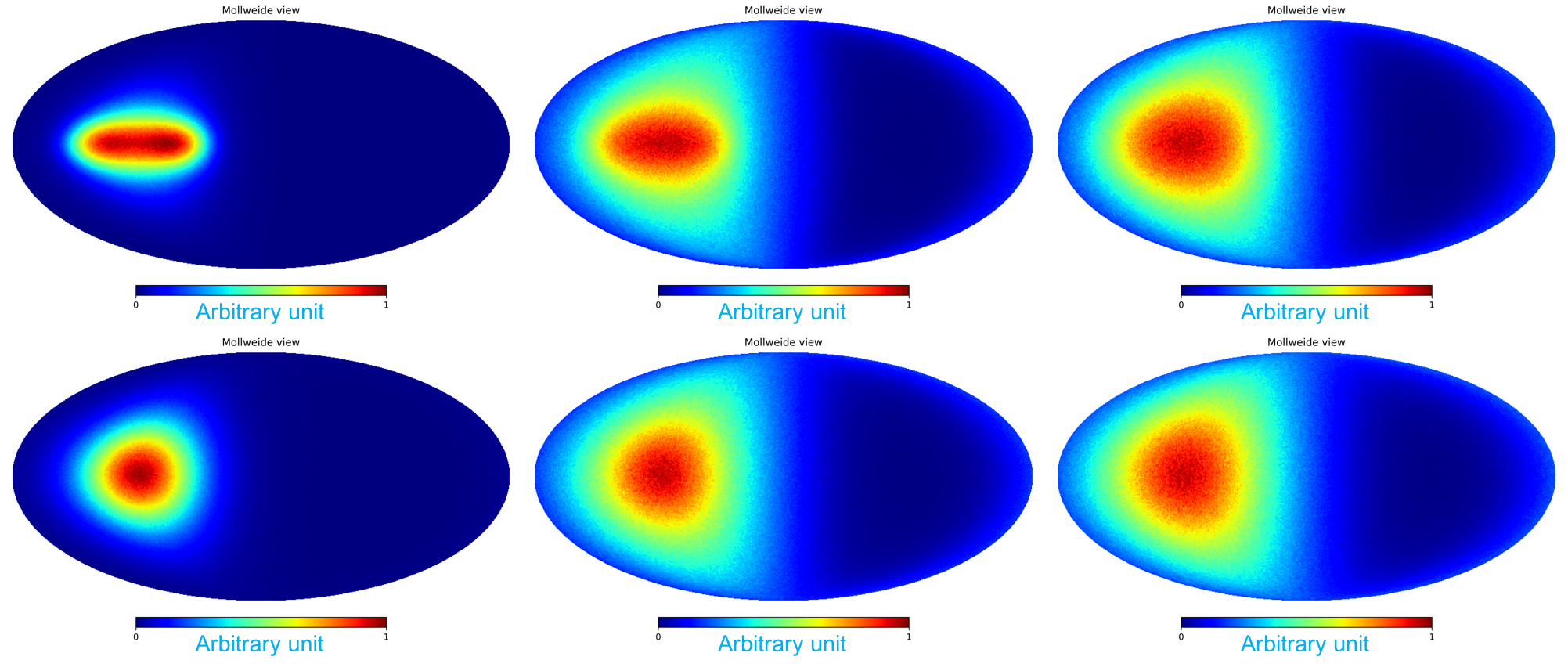}
    \caption{From left to right : WIMP angular distribution, WIMP-induced recoil angular distribution with perfect resolution and with finite resolution ($15^\circ$).
    Top: Pure halo component model with $r=0.0$.
    Bottom: Two-component model with $r=0.5$.}
    \label{fig:recoil-map}
\end{figure}

The WIMP direction distribution and the distribution of the nuclear recoil angle in direct detection with different $r$ value can be found in Ref.~\cite{TaoThese}.
Here without loss of generality, in Figure~\ref{fig:recoil-map} we show the Mollweide projections of the angular distribution of the two-component Galactic WIMP signal and its induced nuclear recoil angular distribution with $r= 0.0$ and $r = 0.5$.
Plots in the middle column are in the case of perfect resolution, while finite resolution of $15^\circ$ are applied in the right column plots.
We conclude that the distribution of the nuclear recoil keeps its dipole feature. 
The impact of the substructure component is faint to be resolved under the current angular resolution.

\section{Conclusion and Outlook}
\label{sec:conclusion}

Providing angular reconstruction with a reasonable angular resolution represents a decisive aspect for Dark Matter identification, and helps overcome the isotropic neutron background and the neutrino floor.
In this paper, we report experiments performed with a MIMAC prototype chamber at the LHI facility with the electric drift field at zero angle with respect to the beam line. 
For $^{19}$F$^{+}$ ion kinetic energies between $6.3$ keV and $26.3$ keV, the angular resolution ranges between $25^{\circ}$ and $3^{\circ}$, respectively. 
We find that down to $10$ keV kinetic energy, the angular resolution ($< 15^{\circ}$) of our MIMAC prototype detector is better than the required $20^{\circ}$ from Ref.~\cite{Billard2011} for small angles.

Systematic effects including diffusion, space-charge at high gain in the amplification gap, and anode pixelization contribute to the measured angular resolution.
Dedicated Garfield++ simulations have shown that these effects allow an experimental access to the angular resolution measurements for small polar angles near $0^\circ$.
The simulations agree with the measurements within $20\%$ for energies above $13.82$ keV. 

Directional detection also plays a key role for neutron spectroscopy from neutron-induced nuclear recoils. 
Several experiments performed with MIMAC detectors with $18$ cm and $25$ cm drift chambers have demonstrated the ability of the MIMAC strategy to perform neutron spectroscopy in the range  $[27\,\mathrm{keV}, 15\,\mathrm{MeV}]$~\cite{Sauzet2019, Tampon2018}. 

When applying the angular resolution we obtain with the MIMAC prototype,  to the two-component Galactic halo model, the directional dipole feature of the WIMP signal is preserved as expected, yet not adequate to resolve the substructure component.

Since we obtain a good angular resolution for sub-millimeter tracks, we can further make use of the favored direction correlated to the latitude of specific underground laboratories (eg. Modane Underground Laboratory (LSM) at $45.18^\circ$N and China Jin-Ping Underground Laboratory (CJPL) at $28.15^\circ$N).

We have shown that the direction of the electric drift field can be a parameter that affects angular resolution,  and therefore it becomes also important to optimize the detector orientation in order to probe the angular distribution of the WIMP-induced recoils in correlation to the rotation of the Earth.

The MIMAC collaboration is developing a 1 m$^3$ detector built from bi-chamber modules with $25$ cm drift each, to be installed in LSM. 
If a Dark Matter signal in the GeV range is observed, the angular resolution that can be obtained with a MIMAC-like detector,  provides a clear signature for the Galactic halo origin of such a Dark Matter candidate.

\section{Acknowledgments}
 
Yi Tao, Igor Moric and Charling Tao thank Tsinghua University physics department and Department of Astronomy (DOA) and National Natural Science Foundation of China (NSFC11475205) for support. 
We acknowledge David Diez for helping the MIMAC team to run Garfield++ simulations.

\bibliographystyle{model1-num-names}
\bibliography{references}

\end{document}